\documentstyle[aps,preprint]{revtex}
\begin{document} \draft

\title {Non-Arrhenius modes in the relaxation of model proteins}

\author{Maksim Skorobogatiy, Hong Guo
and Martin Zuckermann}

\address{Department of Physics and\\
Centre for the Physics of Materials\\
McGill University\\
Montr\'eal, Qu\'ebec, H3A 2T8 Canada.}

\date{\today}
\maketitle

\begin{abstract}

We have investigated the relaxational dynamics for a protein model 
at various temperatures. Theoretical analysis of this model in 
conjunction with numerical simulations suggests several relaxation 
regimes, including a single exponential, a power law and a logarithmic 
time dependence. Even though a stretched exponential form gives a 
good fit to the simulation results in the crossover regime between 
a single exponential and a power law decay, we have not been able 
to directly deduce this form from the theoretical analysis.

\end{abstract}

\vspace{0.5in}


\baselineskip 16pt

\newpage

\section{Introduction}

The manner in which proteins relax to their equilibrium or folded 
configurations is an important subject of current research\cite{dill}. 
In particular the folding kinetics, namely the asymptotics of the
relaxation of a physical quantity such as the total energy, provides
useful clues to the general folding problem. In this work we shall 
provide both numerical and theoretical results concerning the asymptotic 
behavior of relaxation of a model protein following a rapid temperature
quench from high to low. We emphasize and attempt to answer the very 
relevant question:  what is the true (possible) asymptotics of such 
a relaxation of a protein ? Due to the complex nature of this question, 
we shall use a simplified protein model for our numerical simulations
which is based on the monomer-monomer interaction matrix of Ref. \cite{li};
and our analytical work is based on the hierarchically constrained dynamics 
for glassy relaxation\cite{palmer}.

There existed studies of the relaxational dynamics or related equilibrium
properties of protein-like heteropolymers where part or the whole of 
the monomer-monomer interaction is modeled by a stochastic 
term\cite{dill,gutin1,gutin2,parisi,chan,onuchic,guo}. This random 
interaction clearly gives rise to a complicated free energy landscape 
with a large number of highly degenerate energy levels. These heteopolymers 
are believed to simulate certain protein behavior: interesting and important 
results for our intuitive understanding of the folding kinetics have been 
obtained through these studies. Using a lattice copolymer model with Metropolis
Monte Carlo dynamics, Ref. \cite{onuchic} examined the folding kinetics
as a function of the simulation time $t$ for various temperature $T$,
starting from an unfolded initial copolymer configuration. Importantly, 
the folding of the copolymer was found to take place in a two-stage 
process: a rapid collapse followed by a slower adjustment toward the 
ground state, for a broad range of temperatures. A kinetically defined
glass transition temperature $T_g$ and a thermodynamically defined folding
temperature $T_{fold}$ were naturally found from the copolymer simulation 
data\cite{onuchic}. These temperature scales were used to characterize the
foldability of the polymers\cite{onuchic}. Refs.\cite{parisi,guo} directly
measured the relaxation of the total system energy $E$ for a random 
heteropolymer after a temperature quench where the interaction parameters 
between the monomers were drawn from a Gaussian distribution. The Monte 
Carlo\cite{parisi} and molecular dynamics\cite{guo} data were well fit
to a stretched exponential form as a function of the simulation time $t$.
One additional interesting finding of these simulations was that for 
longer heteropolymers the collapse followed a two-stage process\cite{guo}
characterized by different sketching exponents, while for short chains
only one stage was found\cite{parisi}. 

Along another line of development, fruitful results have been obtained from
analyzing even simpler discrete models such as the random energy 
model\cite{derrida}. The application of this model to protein folding
was carried out in Ref. \cite{wolynes} where the folding time was analyzed.
Interestingly, the random energy model gives different relaxation 
asymptotics depending on the details of the transition matrix between 
different microstates\cite{koper} even though the detailed balance condition
has been enforced. Indeed, power-law\cite{koper} decay or
logarithmic\cite{shakhnovich1} decay were both obtained. This is to be
compared with the possible stretched exponential decay\cite{parisi,guo} form.

Previous investigations indicated clearly that a detailed analysis 
of the energy level distribution is one requirement for understanding 
the related relaxation modes. However, this is not sufficient to determine 
the relaxation behavior completely. It is also important to know how the 
random heteropolymer or protein can make transitions between different 
energy levels corresponding to different polymer configurations. If such 
transitions are possible, one says that the levels are connected and the 
numerical values of the transition rates will determine the ease of 
transition. However, due to the microscopic origin of the connectedness 
and the transition rules, it becomes difficult to account for these 
parameters correctly. It is therefore a difficulty problem to predict 
the true asymptotics of a relaxation in these systems.

From the Master Equation approach point of view,  there are difficulties
for the investigation of dynamical systems involving random interactions,
and in particular to predict the precise form of the relaxation process
for folding. Let's consider the relaxation to equilibrium in the 
unbiased random energy model\cite{shakhnovich1}. Here it was believed that 
the relaxation to equilibrium due to a temperature quench from above to below 
the freezing temperature follows a two stage kinetics\cite{shakhnovich1}.
The first stage is characterized by a logarithmic relaxation of the 
extensive part of the energy, while in the second stage, when the system 
has already reached its frozen state, the relaxation follows a power law 
decay. However it has also been suggested that the relaxation dynamics 
can be described by a stretched exponential time dependence\cite{dominicis}, 
or a single power law decay\cite{koper}. In all these references a master 
equation was used to investigate the kinetics of relaxation in the context 
of the random energy model.

Why are there several different predictions for the asymptotics 
of the relaxational dynamics for the same model ?  The answer is related 
to the fact that different transition rates used in the Master Equation 
can lead to different asymptotics\cite{koper}. If we define $W_{ij}$ to 
be a transition rate from the $i$-th energy level to the $j$-th one,
and define $P^{eq}_{i}$, $P^{eq}_{j}$ to be the equilibrium probabilities 
for the two states, then according to detailed balance
$W_{ij}P^{eq}_{i}=W_{ji}P^{eq}_{j}$. In terms of the energies of the 
$i$-th and $j$-th levels, the detailed balance condition takes the form 
$W_{ij}\exp{-\frac{E_{i}}{T}}=W_{ji}\exp{-\frac{E_{j}}{T}}$. Koper and 
Hilhorst suggested\cite{koper} a general form of the transition 
rates: $W_{ij}=W_{o}\exp{-(1-q)\frac{E_{j}}{T}+q\frac{E_{i}}{T}}$ where
$0 \leq q \leq 1$. Adjusting the value of $q$ may lead to different 
asymptotics from the Master Equation.  Indeed, one finds a stretched 
exponential decay\cite{dominicis} by choosing $q=0$, while a power law 
decay\cite{shakhnovich1,koper} for $q=1$ (choosing $q=1$ means that all 
energy levels have the same activation energy).

While the relaxational dynamics by quenching to low temperatures is
clearly complicated, one would in the case of high temperatures  
intuitively expect a single exponential relaxation as can be shown by 
investigating the high temperature limit using a highly connected random 
master equation. At $T \rightarrow \infty$, detailed balance gives 
$W_{ij}=W_{ji}$. Assuming that only a fraction $\gamma$ of states is 
connected, a possible choice of the transition rate is $W_{ij}=1$ with 
probability $\gamma$ and $W_{ij}=0$ with probability $1-\gamma$. A study 
of the related random master equation leads to the conclusion that its 
eigenvalue distribution is sharply peaked at $N \gamma$ with a width of 
the order of $\sqrt{N}$ where $N$ is the number of states. This means that 
there is a dominant time scale which is approximately given by 
$\frac{1}{N \gamma}$, leading to single exponential decay at the high 
temperature region\cite{zwanzig}.

Since the random energy model and a highly connected random master equation 
are related to the problem of protein folding (both approaches are combined 
in the work of Ref. \cite{wolynes}), we arrive at the following picture for 
the behavior of the relaxational dynamics of proteins. At high temperatures, 
relaxation is described by a single exponential decay. Upon quenching from an 
unfolded state to equilibrium with a final temperature of the order of the 
freezing temperature, one might expect a power law, or a stretched 
exponential decay, while quenching to a temperature below the freezing 
point may well lead to an extremely slow logarithmic decay. 

In the following sections we examine the above physical picture carefully 
by means of both numerical and analytical methods. Our analysis also 
provides insight into the crossover in the relaxation from a single 
exponential form at high temperatures to a non-Arrhenius relaxation 
near the freezing temperature. Our numerical simulations are based on Monte
Carlo dynamics\cite{onuchic} for a protein model and we fit the data to
various possible decay modes.  In the theoretical model we note the analogy 
between random heteropolymer relaxation and spin glass dynamics, this allows
us to construct an analytical model for the protein relaxation. The key idea
is that the constraints to the relaxation of the degree of freedom at each
energy level along the relaxation pathway must be taken account\cite{palmer},
and the constraint at a lower energy level depends on what happens on the
higher one.  The asymptotics of our analytical model together with the
simulation data allows a physically reasonable picture to emerge for 
the relaxational dynamics of proteins.

The paper is organized as follows. Section II gives the results of 
numerical simulations for the distinct asymptotic relaxation behavior 
of the folding process at various temperatures using random sequences 
with protein-like energetics\cite{skorobogatiy}. Section III contains our
theoretical analysis where we derive the asymptotics of the relaxation. 
Finally Section IV is reserved for a short summary. The detailed algebra
for the analysis is organized into two Appendices.

\section{Numerical simulation}

We simulated the relaxational dynamics of a model protein by focusing 
on the energy relaxation as a function of the simulation time. The 
simulation was carried out using the Monte-Carlo (MC) method on a 3D 
cubic lattice for a model protein with 27 amino acid residues. The contact 
interaction between the residues is described by the model of Li, Tang 
and Wingreen\cite{li}. In this model, an analysis of the correlations 
between the elements of the Miyazawa-Jernigan (MJ) contact energy 
matrix\cite{m-j}, gave the contact energy of the protein in the following 
form\cite{li,skorobogatiy},
\begin{equation}
E\ =\ \vec{n}\vec{q}\ \ .
\label{eij}
\end{equation}
Here $\vec{n}$ and $\vec{q}$ are $L$ dimensional vectors. $L$ is the 
number of residues: $L=27$ throughout all our numerical simulations.
The vector $\vec{n}$ describes the geometric shape of the protein and 
its element $n_i$ is equal to the number of nearest neighbor contacts 
made by the $i$-th residue with other residues, as discussed in 
Ref. \cite{skorobogatiy}. The vector $\vec{q}$ describes the interaction 
strength of the residues: the $i$-th element of $\vec{q}$ corresponds to 
the strength of the $i$-th monomer on the chain. The values of $q_i$ 
were found from a fit to the MJ matrix in Ref. \cite{li}. Since there are 
twenty different amino acid residues, there will be twenty different values 
of $q_i$ as documented in Ref. \cite{li}. As the MJ matrix was obtained 
from experimental data, the elements of $\vec{q}$ can be viewed as empirical 
parameters to our model. To use Eq. (\ref{eij}), we call two residues 
nearest neighbors if they are not connected along the chain but are 
separated by one lattice constant. In our simulation we fixed the value 
of $q_i$ in the range of $-2.5<q_i<0.0$, where a more hydrophobic amino 
acid residue has a more negative value of $q_i$. For example, for the 
2D model protein depicted in Fig. (\ref{2D-protein}), the first integer 
above each monomer corresponds to the number of nearest neighbors for 
that monomer, while the value $q_i$ of that monomer is written in 
parentheses. Then from Eq. (\ref{eij}), the total energy of this 2D protein is 
$E=3q_1+2q_2 + q_3 + q_4 + q_6 + q_8 + q_9 + q_{11} + q_{12}$. We emphasize 
again that $\vec{n}$ specifies the geometry of the conformation while 
$\vec{q}$ specifies its protein sequence. Finally, we comment that in writing 
down Eq.(\ref{eij}) we have neglected a quadratic 
term\cite{li,skorobogatiy} which gives a small correction to 
the linear term of Eq. (\ref{eij}). However, the relaxational dynamics of 
the model protein is not substantially affected by this quadratic term.

Our simulation considered nine randomly generated protein sequences, 
{\it i.e.} nine different $\vec{q}$ vectors. For each sequence 
we generated a set of relaxation curves corresponding to different 
quenches to different final temperatures and we studied the decay modes by 
averaging over the data for all nine sequences. The allowed moves for a 
monomer in the Monte Carlo simulation were chosen to be the end move, 
the corner move, and the crankshaft move\cite{onuchic}. For a given protein 
sequence, the initial state of the model protein was chosen to be 
completely unfolded so that the initial temperature $T_{i}=16$ was about 
four times higher than that of the folding temperature of our model. After 
the equilibration of the protein at the high temperature $T_{i}$, it was 
quenched to a set of different final temperatures $T_{f}$ and the energy 
relaxation curves were recorded. Every quench $T_{i} \rightarrow T_{f}$ 
consisted of the following operations: a short equilibration of the system 
at high temperature $T_{i}$ using $50,000$ MC steps followed by a quench 
to the final temperature $T_{f}$. The relaxation data was then stored over 
$200,000$ MC steps. After a relaxation curve was recorded the same procedure 
was repeated many times for thermal averages starting from the same 
initial condition: $200$ to $1500$ quenches were averaged depending on 
the value of $T_{f}$ so as to obtain a smooth relaxation curve. Finally, 
we averaged over a set of different initial equilibrium configurations 
each with the same value of $T_i$.

Typical relaxation curves for a particular sequence and a set of different 
$T_{f}$ are presented in Fig. (\ref{rel_curves}). All curves show a fast 
decay immediately after the temperature quench followed by a slow relaxation 
to the asymptotic regime. As we were interested in the non-Arrhenius 
relaxation modes, we fitted these relaxation curves by a stretched 
exponential, a power law and a logarithmic form respectively. The stretched 
exponential decay was taken to have the form 
\begin{equation}
R(t)=a_0+a_1 exp{(-\frac{t}{a_2})^{\gamma}}\ \ .
\label{stretch}
\end{equation}
The power law decay was chosen to be
\begin{equation}
R(t)=a_0+(\frac{a_2}{t})^{\beta}\ \ \ .
\label{power}
\end{equation}
Finally, the logarithmic decay mode was fitted by
\begin{equation}
R(t)=a_0-a_1 ln{t}\ \ \ .
\label{log}
\end{equation}
In equations (\ref{stretch},\ref{power},\ref{log}), $a_0$, $a_1$, 
$a_2$, $\gamma$, and $\beta$ are fitting parameters.
In the following we analyze these fittings.

The fitted value of the stretched exponent $\gamma$ is shown 
in Fig.~(\ref{gamma}) as a function of the final temperature $T_f$. 
This plot is obtained from a particular sequence $\vec{q}$. In the fit we 
used the entire range of the relaxation data where $t \in (2.0,10.0)$ (see
Fig. (\ref{rel_curves})). Although there are substantial error bars for 
$\gamma$, its temperature dependence appears to have three different 
regions. At high $T_f$, the data is consistent with $\gamma\approx 1$. This 
is reasonable since the relaxational dynamics at a high $T_f$ should, as 
discussed above, be a single exponential decay. For an intermediate range 
of $T_f$ as shown in Fig. (\ref{gamma}), $\gamma =\gamma(T_f)$ reduces 
steadily with $T_f$. Finally at very low temperatures, $T_f \in (0.0, 2.0)$, 
$\gamma$ has a value close to $0.3$. However, such a small value of 
$\gamma$ usually indicates that the decay is actually not in a 
stretched exponential form but has instead a different relaxational 
behavior such as a power law.

The result of fitting the relaxation curves to a power law is summarized 
in Fig. (\ref{beta1}), where the fitted power $\beta$ is shown as a 
function of $T_f$. The difficulty in fitting to a power law is that a 
divergence occurs at the quenching time $t=2$ (see Fig. (\ref{rel_curves})).
Since we expect the power law decay to be a possible asymptotic
of the relaxation, we used the simulation data for the time period of 
$t \in (2.1,10.0)$ to avoid the divergence. Also, as the final temperature 
$T_f$ increases, the characteristic relaxation time (parameter $a_2$) 
decreases substantially and it becomes quite difficult to fit the data 
to a power law accurately. This is the reason why the dependence of 
$\beta$ on $T_f$ shown on Fig. (\ref{beta1}) is only shown below $T_{f}=3.0$. 
Nevertheless, it is clear that this dependence in the range
$T_{f} \in (2.0,3.0)$ is given by roughly a straight line, namely 
$\beta \sim T_f-T^{\beta}_{f}$. Extrapolating this line backwards gives 
$T^{\beta}_{f} \sim 1.5$. An important feature of Fig. (\ref{beta1}) is that 
as $T_f$ is in the range $\in (0.0,2.0)$, the relaxation become extremely 
slow and the power $\beta$ levels off at a small value $\beta \sim 0.3$. 
The fact that both the stretched exponent $\gamma$ and the power law 
$\beta$ change their behavior at nearly the same value of $T_f (\sim 2.0$) 
signals that a transition to a frozen state occurs near this temperature. 

We attempted to fit the relaxation data for the low temperature range 
$T_f \in (0.0,2.0)$ with a logarithmic decay. An almost perfect fit was 
found for all the sequences at $T^{log}_{f} \sim 1.5$, as shown in 
Fig. (\ref{log1}). Note that $T^{log}_{f} \sim T^{\beta}_{f}$. 
We believe that the logarithmic decay mode sets in near the freezing
temperature as demonstrated by the random energy model\cite{shakhnovich1}.
We thus denote this special temperature scale to be $T_G$:
$T_G\sim T^{log}_{f} \sim T^{\beta}_{f}$. We shall use the notation $T_G$ 
in the rest of this paper as it is reminiscent of a glass transition 
temperature. Note that the nature of glass transition temperature
for copolymer systems has been thoroughly examined in Ref. \cite{onuchic} and
we refer interested readers to that article for details concerning the 
related physics.

Our numerical prediction of the occurrence of a low temperature 
logarithmic decay mode at temperature $\sim T_G$ agrees with the 
prediction for the random energy model of Ref. \cite{shakhnovich1}. In 
addition, from the theoretical analysis of the random energy 
model in Ref. \cite{koper}, it follows that the relaxation near 
$T_G$ should follow a power law decay with a power $\beta\sim (T-T_G)$. 
Our simulation data presented in Fig. (\ref{beta1}) is consistent with 
this linear dependence. On the other hand, we found that fitting the data 
near the freezing point $T_G$ with a stretched exponential does not 
produce an exponent $\gamma$ lying along the line which would be expected 
from Ref. \cite{dominicis}. Finally, from our numerical data it still 
remains unclear whether the asymptotics {\it above} the freezing temperature 
follows a stretched exponential decay, although it is apparent that the
data can be reasonably well fitted by such a stretched exponential. 

\section{Theoretical analysis}

To understand the simulation data presented in the last section, in the
following we make a theoretical analysis of the relaxation based on the 
model of hierarchically constrained dynamics for glassy system
relaxation\cite{palmer}. 

The conventional Arrhenius relaxation is described by a single exponential 
decay with a time scale $\tau$ 
\begin{equation}
R(t)\ =\ R_0 exp{(-\frac{t}{\tau})}
\end{equation}
where $R(t)$ is the physical quantity of interest. We obtain a more 
complicated relaxation behavior by assuming a distribution of time scales. 
Suppose that $P(\tau)$ is a continuous distribution of possible time 
scales in our system, then, using the notion of parallel relaxation we write 
for the total relaxation process\cite{palmer} 
\begin{equation}
R(t)=R_0 \int^{\tau_{max}}_{\tau_{min}} P(\tau)
exp{(-\frac{t}{\tau})}d \tau \ \ \ .
\label{rel1}
\end{equation}
This approach is especially attractive in the case where dynamics is 
modeled by the Master Equation. As the Master Equation is a first order 
differential equation, the general solution will be a sum of exponential 
decays with time scales inversely proportion to the eigenvalues $\lambda$
of the transition matrix. A distribution of time scales $P(\tau)$ thus 
corresponds to the distribution of inverse eigenvalues $P(\frac{1}{\lambda})$.

As pointed out by Palmer {\it et. al.}\cite{palmer}, though a description 
of relaxation by (\ref{rel1}) is certainly appealing, the nature of 
$P(\tau)$ is not {\it a priori} clear. In the dynamics of such complex 
systems as random polymers or proteins, it is intuitively plausible that 
$P(\tau)$ should take into account the dynamic constraints: {\it e.g.} 
monomer $A$ cannot move until monomer $B$ moves out of the way. $P(\tau )$ 
may also depend on such factors as ergodicity breaking in a frozen state 
of the system. The theory should also involve a hierarchy of degrees of 
freedom from fast to slow. For the protein problem, the fastest degrees of 
freedom could involve single-atom motion, while slower degrees of freedom 
could involve a diffusion of domains of secondary structure.

The original theory of Ref. \cite{palmer} considered a discrete set of 
levels $n=1,2,3...,N$ with the degrees of freedom in level $n$ represented 
by $N_n$ pseudospins $S_i$. Each spin in the $(n-1)$th level is only free 
to change its state if a condition on some spins at level $n$ is satisfied, 
thus providing a hierarchy of constraints. This is the key idea of the 
model as shown schematically in Fig. (\ref{hiera}). We take the constraint 
to be that $\tilde{\mu}_n$ spins in level $n$ ($\tilde{\mu}_n < N_n$) 
attain one particular state out of their $2^{\tilde{\mu}_n}$ possible states. 
Then the average relaxation times $\tau_{n-1}$ will be related to $\tau_{n}$ 
as $\tau_{n-1}=2^{\tilde{\mu}_n}\tau_{n}$.  Hence, 
\begin{equation}
\tau_{n}=\tau_{N}exp(\sum^{N}_{i=n} \mu_i)
\label{rel2}
\end{equation}
where $\mu_i=\tilde{\mu}_i ln{2}$, and $\tau_N$ sets the time scale of the 
relaxation. The average relaxation can now be written as follows
\begin{equation}
R(t)=R_0 \sum^{N}_{n=1} N_n exp{(-\frac{t}{\tau_n})}
\label{rt}
\end{equation}
\begin{equation}
\tau_{n}=\tau_{N}exp(\sum^{N}_{i=n} {\mu}_i)\ \ .
\end{equation}

To apply this theory to protein folding dynamics, we notice that the 
summation is taken over $(N-n)$ higher energy levels in Eq. (\ref{rel2}) 
since the states corresponding to the higher energy levels (the 
unfolded states) relax very fast because of minimal dynamical constraints. 
Clearly, as the folding proceeds and the level index $n$ decreases, 
there will be an increased number of dynamical constraints on folding as the
protein becomes geometrically more compact. Hence the relaxation times for 
the more folded states should increase in comparison to those of the 
less folded states. The meaning of the ``pseudospins'' of Ref. \cite{palmer} 
in our protein problem can be elucidated by noticing the similarity 
of a spin flip and a local monomer move. For instance, a crankshaft move, 
a corner move or an end move in our Monte Carlo simulation could 
be compared with a pseudospin flip,  while a configuration of spins could 
represent a protein conformation. Two energetically close protein 
conformations may differ from each other by a local move of a monomer, 
similar to the flip of a spin. It is much easier to make a local 
move (a spin flip) in an unfolded state than in a folded state because of the 
energetics consideration and the importance of volume exclusion interactions 
in compact protein conformations. Clearly, the analogy of pseudospin flips 
to local monomer moves can be extended to more realistic off-lattice protein 
models.

From the above discussion, we now make some assumptions about the 
form of $\mu_i$ and $N_n$ for the protein problem. First, from the form 
of Eq. (\ref{rel2}) one can immediately deduce that $\mu_i$ should depend 
upon temperature. When $T\rightarrow \infty$, the relaxation should be a 
single exponential, from Eq. (\ref{rt}) we conclude 
$\tau_n (T \rightarrow \infty)=const$ and it should not depend on $n$. 
The simplest form for $\mu_i$ satisfying these conditions is 
$\mu_i=\frac{T^{f}_i}{T}$. As far as $N_n$ is concerned, it should be 
proportional to the total number of states on each level. 
When $T\rightarrow \infty$, all the pseudospins are excited so that $N_n$ 
in this limit equals to the number of pseudospins on the $n$th level. 
As temperature decreases and the system finally goes into the 
frozen state, the number of pseudospins on each level should decrease to 
the order of unity signifying ergodicity breaking in the relaxational 
dynamics. For our protein problem, in the low temperature limit the 
decrease in the number of pseudospins corresponds to the restriction of 
the dynamical space to a set of specific pathways for protein relaxation. 
In our model we choose $N_n$ to depend on temperature in the form
\begin{equation}
N_n=\lambda^{n(1-\mu_{g})}
\label{nn}
\end{equation}
where $\mu_{g}=\frac{T^{g}_n}{T}$ and $T^{g}_n$ corresponds to the 
``level freezing" temperature of Ref. \cite{onuchik}. Summarizing all 
these considerations, our complete model for protein relaxational
dynamics is given by
\begin{equation} 
R(t)=R_0 \sum^{N}_{n=1} \lambda^{n(1-\frac{T^{g}_n}{T})}
exp{(-\frac{t}{\tau_n})}
\end{equation}
\begin{equation}
\tau_{n}=\tau_{N}exp(\sum^{N}_{i=n} \frac{T^{f}_i}{T})\ \ .
\label{taun}
\end{equation}
It is convenient to write our model in an integral form, so that 
\begin{equation}
R(t)=R_0 \int^{N}_{1} N(n) \exp{(-\frac{t}{\tau_{n}})} dn \ \ \ .
\end{equation}
Here the function $N(n)$ is the continuous version of (\ref{nn}).

To make further analysis tractable and considering the fact that $N$ 
is large, we make a reasonable assumption that $T^{g}_n$ and $T^{f}_n$ 
to be roughly constants for all energy levels, so that 
$\mu_{g}=\frac{T^{g}}{T}$ and $\mu_{f}=\frac{T^{f}}{T}$. The above 
expressions can be written in terms of the distribution of time 
scales $\tau$: 
\begin{equation}
R(t)=R_0 \int^{\tau_{max}}_{\tau_{min}} (N(n)
\frac{dn}{d \tau})|_{\tau=\tau_n} \exp{(-\frac{t}{\tau})} d \tau 
\label{rt3}
\end{equation}
where $\tau_{min}=\tau_N$ and 
$\tau_{max}=\tau_{N} \exp{(\sum^{N}_{i=1} \mu_i)}$.  Comparing the 
above equation with (\ref{rel1}), we notice that the distribution of 
time scales is just
\begin{equation}
P(\tau)=N(n) \frac{dn}{d \tau}|_{\tau=\tau_{n}}\ \ ,
\end{equation}
where the right hand side is understood as replacing all $n$ dependence by
$\tau$ through the transformation $\tau=\tau_n$ using Eq. (\ref{taun}).
It is then straightforward to obtain,
\begin{equation}
R(t)=R_0 \int^{\tau_{max}}_{\tau_{min}} P(\tau)
exp{(-\frac{t}{\tau})}d \tau
\end{equation}
with an unnormalized distribution of time scales
\begin{equation}
P(\tau)\sim (\frac{\tau_N}{\tau})
^{\frac{(1-\mu_{g})ln{\lambda}}{\mu_f}+1}\ \ .
\label{ptau}
\end{equation}
Denoting 
\begin{equation}
\beta\equiv \frac{1-\mu_{g}}{\mu_f}ln{\lambda}
=\frac{T-T^{g}}{T^{f}}ln{\lambda}
\label{beta}
\end{equation}
we obtain a main result of this analysis, 
\begin{equation}
R(t)=R_0 \int^{\tau_{max}}_{\tau_{min}}
(\frac{\tau_N}{\tau})^{\beta+1}
\exp{(-\frac{t}{\tau})} d \tau\ \ .
\label{result1}
\end{equation}
One of the important outcomes of our model is that the distribution 
of time scales is a power law. We now discuss several possible cases 
for different temperatures.

\subsection{The case of $T<T^{g}$}

For low temperatures $T < T^g$, from Eq. (\ref{beta}) $\beta<0$. We define 
$\eta\equiv -\beta$ so that $\eta>0$. Then the normalized time scale 
distribution (\ref{ptau}) becomes 
\begin{equation}
P(\tau)=\frac{\eta}{\tau_{min}\left[\left(
\frac{\tau_{max}}{\tau_{min}}\right)^{\eta}-1\right]}
\left(\frac{\tau_{min}}{\tau}\right)^{1-\eta}\ \ .
\end{equation}
The relaxation, obtained from Eq. (\ref{result1}), is
\begin{equation}
R(t)=\frac{\eta t^{\eta}}{\tau_{max}^{\eta}-\tau_{min}^{\eta}}
\int^{\frac{t}{\tau_{min}}}_{\frac{t}{\tau_{max}}}
\chi^{-1-\eta}exp{(-\chi)}d \chi
\end{equation}
where $\tau_{min}=\tau_N$ and $\tau_{max}=\tau_N exp{(N\frac{T^f}{T})}$. 
$R(t)$ is normalized so that $R(t)|_{t=0}=1.0$. Notice that we cannot 
simply extend the integration from $(\tau_{min},\tau_{max})$ to 
$(0,\infty)$ because the integral would then diverge. Still, it is not 
difficult to obtain an asymptotic form of this expression as 
$\eta \rightarrow 0$ and $t \gg \tau_{min}$. From the detailed analysis 
presented in Appendix I, the final asymptotic of the relaxation for
this temperature range is given by,
\begin{equation}
R(t) \asymp c_1-c_2 ln{t}
\end{equation}
where $c_1$ and $c_2$ are constants. 

Fitting the low temperature simulation data of the last section to this 
logarithmic decay, as shown in Fig. (\ref{log1}), a very good fit is 
obtained in the interval $1.5 \ge T \ge 1.25$. Since $T^g$ is the level 
freezing temperature, and when all the individual energy levels freeze, 
the whole system freezes. Thus we must have $T^g\sim T_G$ which is the 
freezing transition temperature discussed in section II. This allows us
to conclude that the hierarchically constrained dynamics model presented
in this section has a logarithmic decay mode for low temperatures $T<T_G$
as the true asymptotic.

\subsection{The case of $T>T^{g}$}

When at higher temperatures $T>T^{g}$, from Eq. (\ref{beta}) we have 
$\beta>0$. In this case, a normalized time scale distribution, (\ref{ptau}), 
will be
\begin{equation}
P(\tau)=\frac{\beta}{\tau_{min}
\left[1-\left(\frac{\tau_{min}}{\tau_{max}}\right)^{\beta}\right]}
\left(\frac{\tau_{min}}{\tau}\right)^{1+\beta}\ \ .
\end{equation}
The relaxation is thus
\begin{equation}
R(t)=\frac{\beta}{(\frac{1}{\tau_{min}})^
{\beta}-(\frac{1}{\tau_{max}})^{\beta}}\frac{1}
{t^{\beta}}
\int^{\frac{t}{\tau_{min}}}_{\frac{t}{\tau_{max}}}
\chi^{\beta-1}exp{(-\chi)}d \chi \ \ .
\label{rt1}
\end{equation}
As $\tau_{max} \gg t \gg \tau_{min}$, the integral in $R(t)$ can be 
extended from $(\tau_{min},\tau_{max})$ to $(0,\infty)$ without
introducing too much error (the integral is also convergent in the 
whole interval). Hence,
\begin{equation}
R(t) \asymp const(\beta) (\frac{\tau_{min}}{t})^{\beta}\ \ .
\end{equation}
We thus obtain a power law decay for the relaxation when $T>T^g$. 
Moreover, since the exponent $\beta=\frac{T-T^{g}}{T^{f}}ln{\lambda}$, 
we can expect a linear dependence of $\beta$ as a function of temperature 
near the freezing transition point. Our simulation data of Fig. (\ref{beta1}) 
agrees with this result and gives $T^{g}=1.5$, 
$\frac{T^{f}}{ln{\lambda}}=1.2$\ .

\subsection{The case of $T \gg T^{g}$}

We note that formally, the case of $\beta>0$ should be 
divided into two regimes: $1>\beta>0$ and $\beta \ge 1$. 
The latter has a special point $\beta=1$ because of the 
integrand of Eq. (\ref{rt1}) which can be written as follows
\begin{equation}
\Theta(\beta,t)\equiv \int^{\frac{t}{\tau_{min}}}_{0}
\chi^{\beta-1}exp{(-\chi)}d \chi \ \ .
\end{equation}
Taking into account the continuous behavior of $\Theta(\beta,t)$ 
for $\beta>1$, let us calculate the relaxation
\begin{equation}
R(t)=\frac{\beta}{(\frac{1}{\tau_{min}})^
{\beta}-(\frac{1}{\tau_{max}})^
{\beta}}\frac{1}{t^{\beta}}
\Theta(\beta,t)
\end{equation}
as $\beta=p$ where $p$ is any positive integer. In this case 
$\Theta(\beta,t)$ can be calculated analytically and the 
details are given in Appendix II. The relaxation function 
is found to be 
\begin{equation}
R(t)=exp{(-\frac{t}{\tau_{min}})}
(1+\frac{(\frac{t}{\tau_{min}})}{(p+1)}+
\frac{(\frac{t}{\tau_{min}})^2}
{(p+1)(p+2)}+\frac{(\frac{t}
{\tau_{min}})^3}{(p+1)(p+2)(p+3)}+...)
\end{equation}
Hence, as temperature increases, non-Arrhenius decay modes 
would be observed as asymptotics at $t \gg (p+1)\tau_{min}$, while 
an initial stage of relaxation $t \lesssim (p+1)\tau_{min}$ will be an 
Arrhenius-like single exponential decay. As temperature 
increases, the overall relaxation time to equilibrium 
becomes shorter and, finally, it becomes impossible to 
distinguish non-Arrhenius asymptotics at the tail of the relaxation 
function. Notice that the onset of the single exponential decay occurs 
at temperatures corresponding to $\beta \gtrsim 1$ or
$T \gtrsim T^{g}+\frac{T^{f}}{ln{\lambda}}$. From the 
simulation data of Fig. (\ref{beta1}), we can calculate 
that this occurs as $T \gtrsim 2.7$ which corresponds 
approximately to the midpoint of the crossover regime 
from $\gamma=1.0$ to $\gamma \sim 0.3$ on Fig. (\ref{gamma}).  This 
result indicates that although one can {\it fit} the simulation data as a 
stretched exponential, it is quite possible that this fit does not in 
fact give the correct asymptotic nature of the crossover from a single 
exponential to a power law decay as temperature changes.

Finally, we comment that we were not able to obtain a stretched 
exponential asymptotic decay from our analytical theory. In general, a 
transition from single exponential decay at high temperatures to power 
law decay at lower temperatures could proceed via a set of more intricate 
functions reminiscent of stretched exponentials. However, 
whether these functions have stretched exponentials as 
their true asymptotics can not as yet be resolved from the theory
presented in this section.

\section{Summary}

In this work we addressed the problem of the relaxational dynamics of a 
model protein for temperature quenches from an unfolded state at high 
temperature to a range of low temperatures. The results of our Monte Carlo
simulations and our theoretical analysis of a hierarchical model 
proposed for the protein relaxation dynamics clearly indicate three 
different decay modes: first a logarithmic decay at $T\lesssim T^{g}$, 
next an asymptotic power law decay at $T \gtrsim T^{g}$ with the power 
$\beta \sim (T-T^{g})$ and finally, as the temperature increases to $T 
\gtrsim T^{g}+\frac{T^{f}}{ln{\lambda}}$, a single exponential decay. 
Here the temperature scale $T^g$ describes a freezing transition where 
the decay mode changes. Our numerical and analytical results suggest that 
a crossover from a power law relaxation to a single exponential decay 
occurs via an intricate interplay between the time intervals for which 
these decay modes are valid, although this crossover can be reasonably 
fitted numerically to a stretched exponential form.

\section*{Acknowledgments}

The authors wish to thank Christine Villeneuve for helpful discussions
and assistance concerning the Monte Carlo simulations. We gratefully 
acknowledge support by the Natural Sciences and Engineering Research 
Council of Canada and le Fonds FCAR du Qu\'ebec. One of the authors 
(MJZ) is an Associate of the Canadian Institute for Advanced Research.

\section*{Appendix I}

In the appendix we derive the logarithmic decay form for low temperatures
$T<T_g$. As stated in section III, the relaxation function at $\beta<0$, 
$\eta=-\beta$ is 
\begin{equation}
R(t)=\frac{\eta
t^{\eta}}{\tau_{max}^{\eta}-\tau_{min}^{\eta}}
\int^{\frac{t}{\tau_{min}}}_{\frac{t}{\tau_{max}}}
\chi^{-1-\eta}exp{(-\chi)}d \chi\ \ .
\end{equation}
We can rewrite this expression in a form
\begin{equation}
R(t)=\frac{\eta
t^{\eta}}{\tau_{max}^{\eta}-\tau_{min}^{\eta}}
\int^{\frac{t}{\tau_{min}}}_{\frac{t}{\tau_{max}}}
\chi^{-1-\eta}(\sum_{i=0}^{\infty}
\frac{(-1)^i }{i!} \chi^{i}) d \chi\ \ .
\end{equation}
After integration of the $\chi$ variable, we have
\begin{equation}
R(t)=1+\frac{\eta}{\tau_{max}^{\eta}-\tau_{min}^
{\eta}} \left[\sum_{i=1}^{\infty} \frac{(-1)^i t^i}{i!(i-\eta)}
\left(\frac{1}{\tau_{min}^{i-\eta}}-\frac{1}{\tau_{max}^{i-\eta}}\right)
\right]\ \ .
\end{equation}
As $\eta \rightarrow 0$ and noticing $\tau_{max} \gg \tau_{min}$, 
$R(t)$ becomes
\begin{equation}
R(t)=1+\frac{1}{ln{\frac{\tau_{max}}{\tau_{min}}}}
\left[\sum_{i=1}^{\infty} \frac{(-1)^i}{i!i}
\left(\frac{t}{\tau_{min}}\right)^i\right]\ +\ O(\eta) \ \ .
\end{equation}
To sum up the series, we differentiate R(t) to obtain
\begin{equation}
R(t)'=\frac{1}{ln\left(\frac{\tau_{max}}{\tau_{min}}\right)\tau_{min}}
\left[\sum_{i=1}^{\infty} \frac{(-1)^i}{i!}
(\frac{t}{\tau_{min}})^{i-1}\right]\ +\ O(\eta) \ \ .
\end{equation}
We can now perform the summation to obtain
\begin{equation}
R(t)'=\frac{1}{ln\left(\frac{\tau_{max}}{\tau_{min}}\right)t}
\left[e^{-\frac{t}{\tau_{min}}}-1\right]\ +\ O(\eta)\ \ .
\end{equation}
As $t \gg \tau_{min}$, $R(t)'$ becomes even simpler
\begin{equation}
R(t)'=-\frac{1}{ln\left(\frac{\tau_{max}}{\tau_{min}}\right)t}\ +\ O(\eta) 
\ .
\end{equation}
Final integration yields
\begin{equation}
R(t)=const-\frac{1}{ln\left(\frac{\tau_{max}}{\tau_{min}}\right)}ln{(t)}
\ +\ O(\eta) 
\end{equation}
which is the logarithmic decay.

\section*{Appendix II}

For $\beta>0$,
\begin{equation}
\Theta(\beta,t)=\int^{\frac{t}{\tau_{min}}}_{0}
\chi^{\beta-1}exp{-\chi}d \chi   
\end{equation}
The relaxation function is then given by
\begin{equation}
R(t)=3D\frac{\beta}{(\frac{1}{\tau_{min}})^{\beta}-
(\frac{1}{\tau_{max}})^{\beta}}\frac{1}{t^{\beta}}
\Theta(\beta,t)\ \ .
\end{equation}
Choosing $\beta=p$, where $p$ is a positive
integer, and assuming that $\tau_{max} \gg \tau_{min}$, we obtain
\begin{equation}
\Theta(\beta,t)=\int^{\frac{t}{\tau_{min}}}_{0}
\chi^{p-1}exp{(-\chi)}d \chi 
\end{equation}
and
\begin{equation}
R(t)=p(\frac{\tau_{min}}{t})^p\Theta(p,t)\ \ .
\label{rel10}
\end{equation}
$\Theta(\beta,t)$ is a standard integral and can
be evaluated to give 
\begin{equation}
\Theta(\beta,t)=3D(p-1)!(1-exp{(-\frac{t}{\tau_{min}})}
(1+\frac{(\frac{t}{\tau_{min}})}{1!}+
\frac{(\frac{t}{\tau_{min}})^2}{2!}+...+\frac
{(\frac{t}{\tau_{min}})^{p-1}}{(p-1)!}))\ \ .
\label{rel11}
\end{equation}
Noticing that
\begin{equation}
1+\frac{(\frac{t}{\tau_{min}})}{1!}+
\frac{(\frac{t}{\tau_{min}})^2}{2!}+...+\frac
{(\frac{t}{\tau_{min}})^{p-1}}{(p-1)!}=
exp{(\frac{t}{\tau_{min}})}-\sum^{\infty}_{p}
\frac{(\frac{t}{\tau_{min}})^i}{i!}
\end{equation}
after substitution to (\ref{rel11}) and (\ref{rel10}) we find
\begin{equation}
R(t)=exp{(-\frac{t}{\tau_{min}})}(1+\frac{(\frac{t}
{\tau_{min}})}{(p+1)}+
\frac{(\frac{t}{\tau_{min}})^2}{(p+1)(p+2)}+\frac{
(\frac{t}{\tau_{min}})^3}{(p+1)(p+2)(p+3)}+...)
\end{equation}
which is the desired result.

\begin{figure}
\caption{A 2D lattice model of a protein. The figure over each
monomer indicates the number of its nearest neighbors. The parameter in 
parentheses indicates the strength of amino-acid situated on a particular site.
}
\label{2D-protein}
\end{figure}

\begin{figure}
\caption{A typical set of relaxation curves for a temperature quench
$T_{i} \rightarrow T_{f}$, for a given sequence and different 
final temperatures.  The unit of time is $25,000$ Monte Carlo steps.
}
\label{rel_curves}
\end{figure}

\begin{figure}
\caption{Fit to a stretched exponential decay. All nine $\gamma-T_f$ 
curves for different sequences were averaged to obtain the 
averaged behavior of the stretched exponential relaxation 
as a function of the final temperature of the quench. An individual 
$\gamma-T_f$ curve for a particular sequence was obtained 
by fitting the relaxation curves to a stretched exponential form. 
}
\label{gamma}
\end{figure}

\begin{figure}
\caption{Fit to a power law decay. All nine $\beta-T_f$ 
curves for different sequences were averaged to obtain the
average behavior of the power law relaxation as a 
function of the final temperature of the quench. An individual 
$\beta-T_f$ curve for a particular sequence was 
obtained by fitting the relaxation curves to a power law 
decay form. 
}
\label{beta1} 
\end{figure}

\begin{figure}
\caption{Fit to a logarithmic decay mode for a specific sequence.
Similar fits for other sequences show that the best fit 
is found for $T_{f} \sim 1.5$.
\label{log1}
}
\end{figure}

\begin{figure}
\caption{Model of hierarchically constrained dynamics for 
relaxation. The probability of the pseudospin flip on 
the $(n-1)th$ level is coupled to the probability of occurrence of a 
specific configuration of pseudospins on level $n$. 
\label{hiera}} 
\end{figure} 

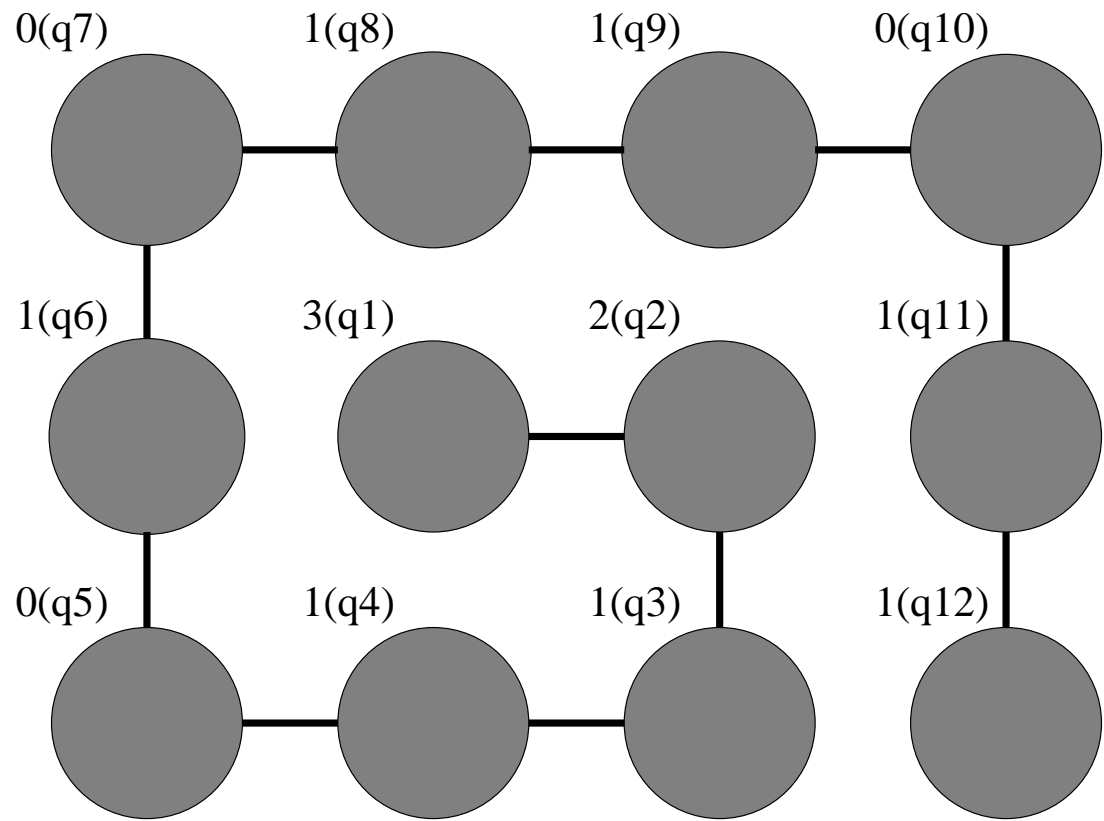
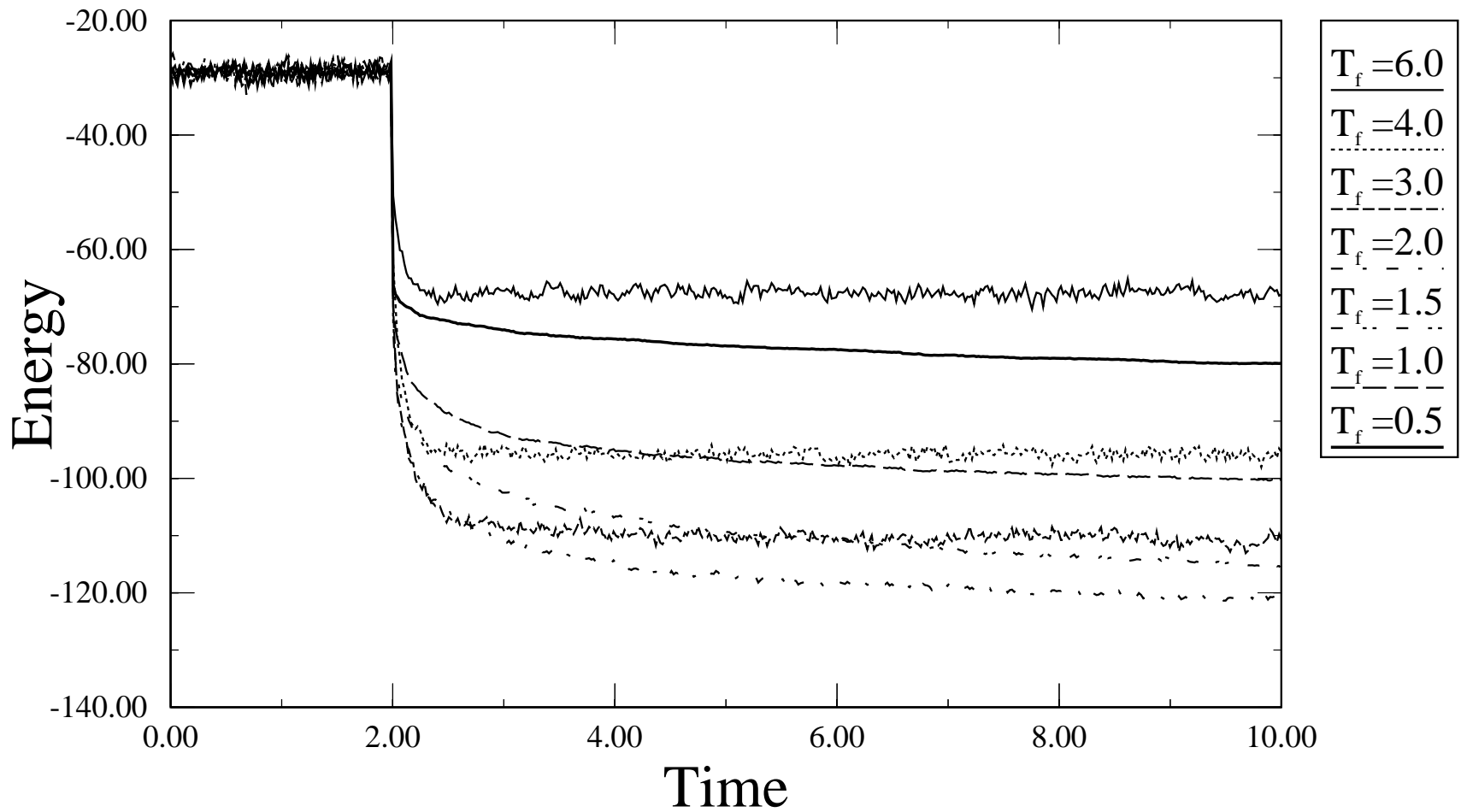
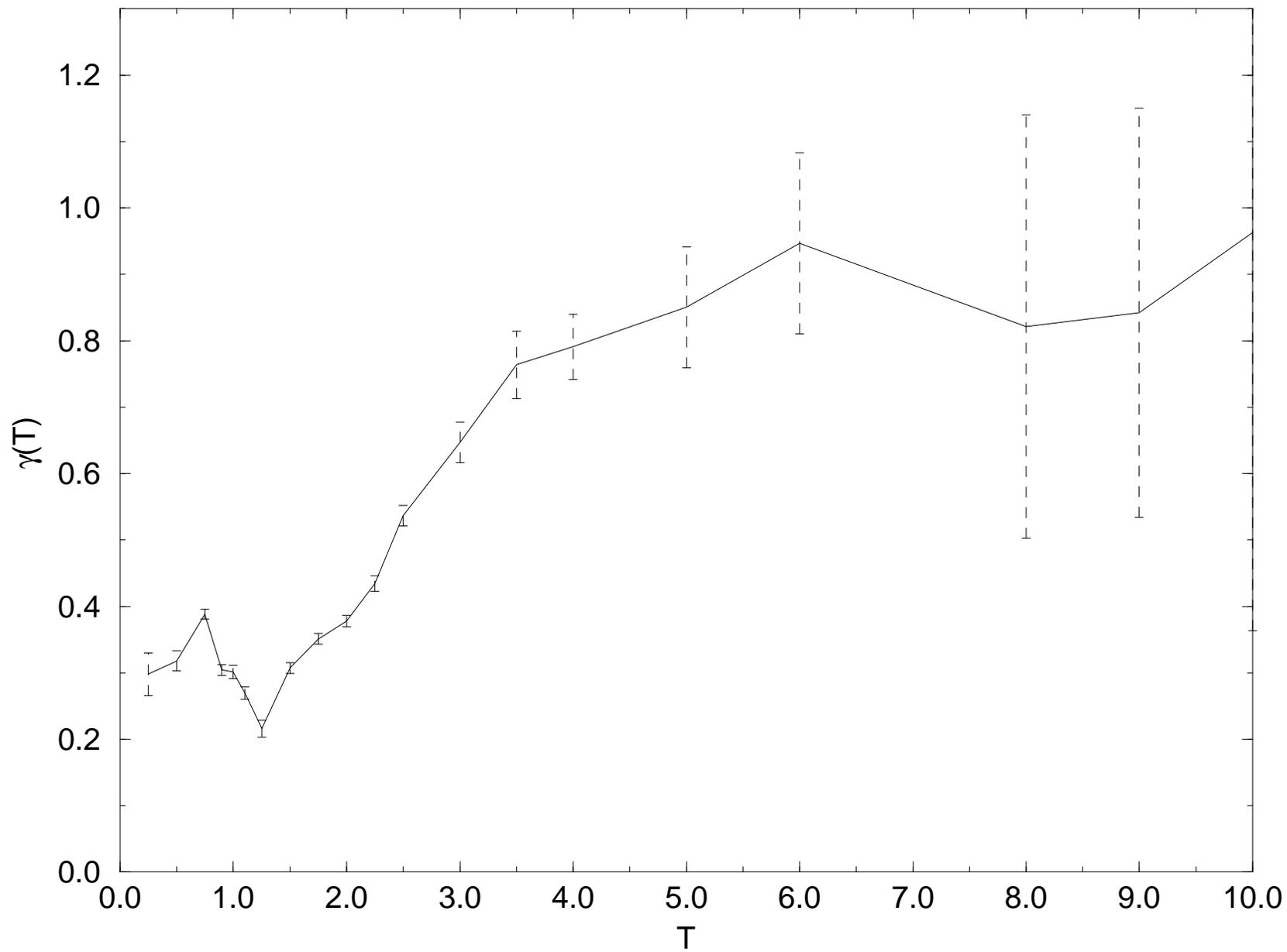
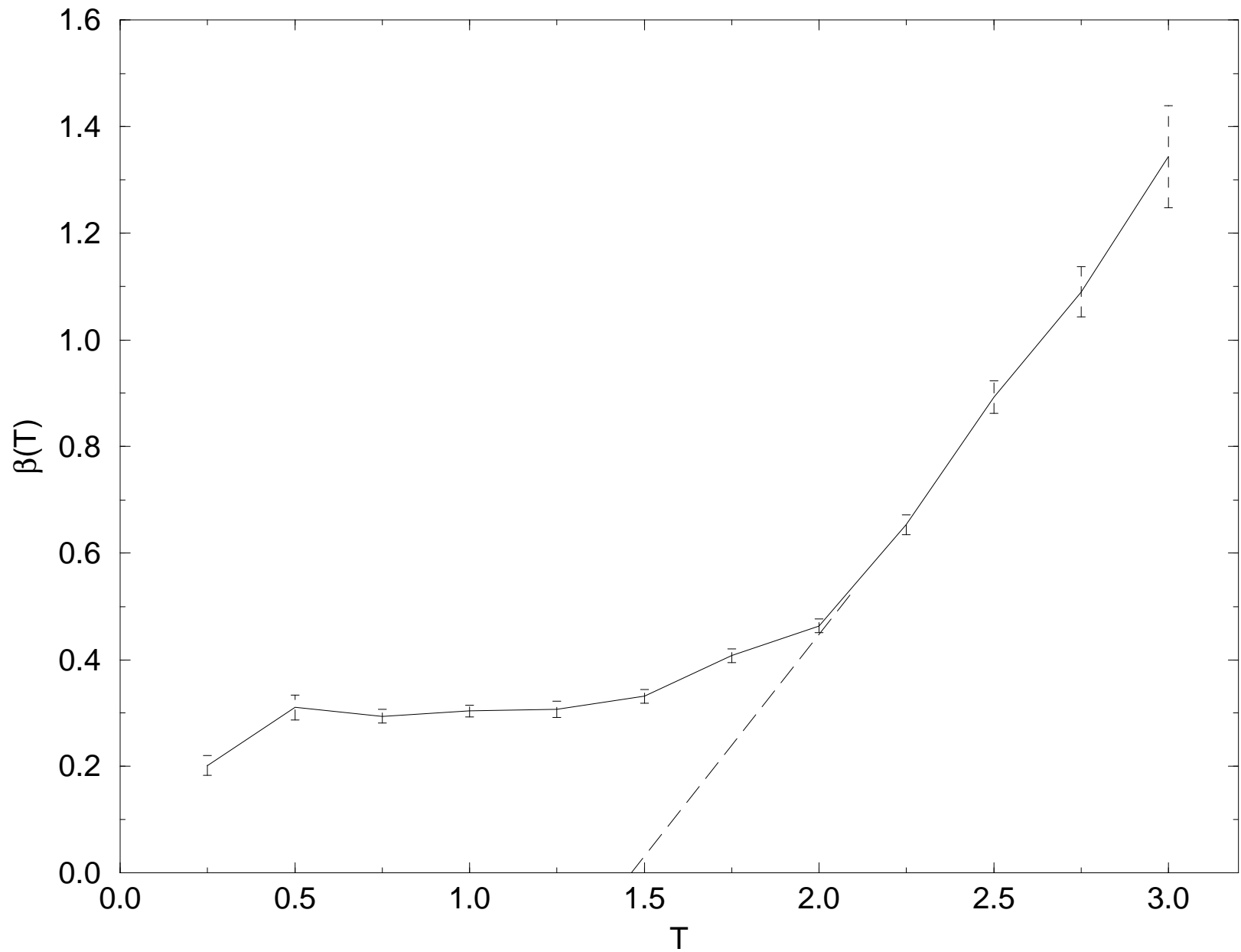
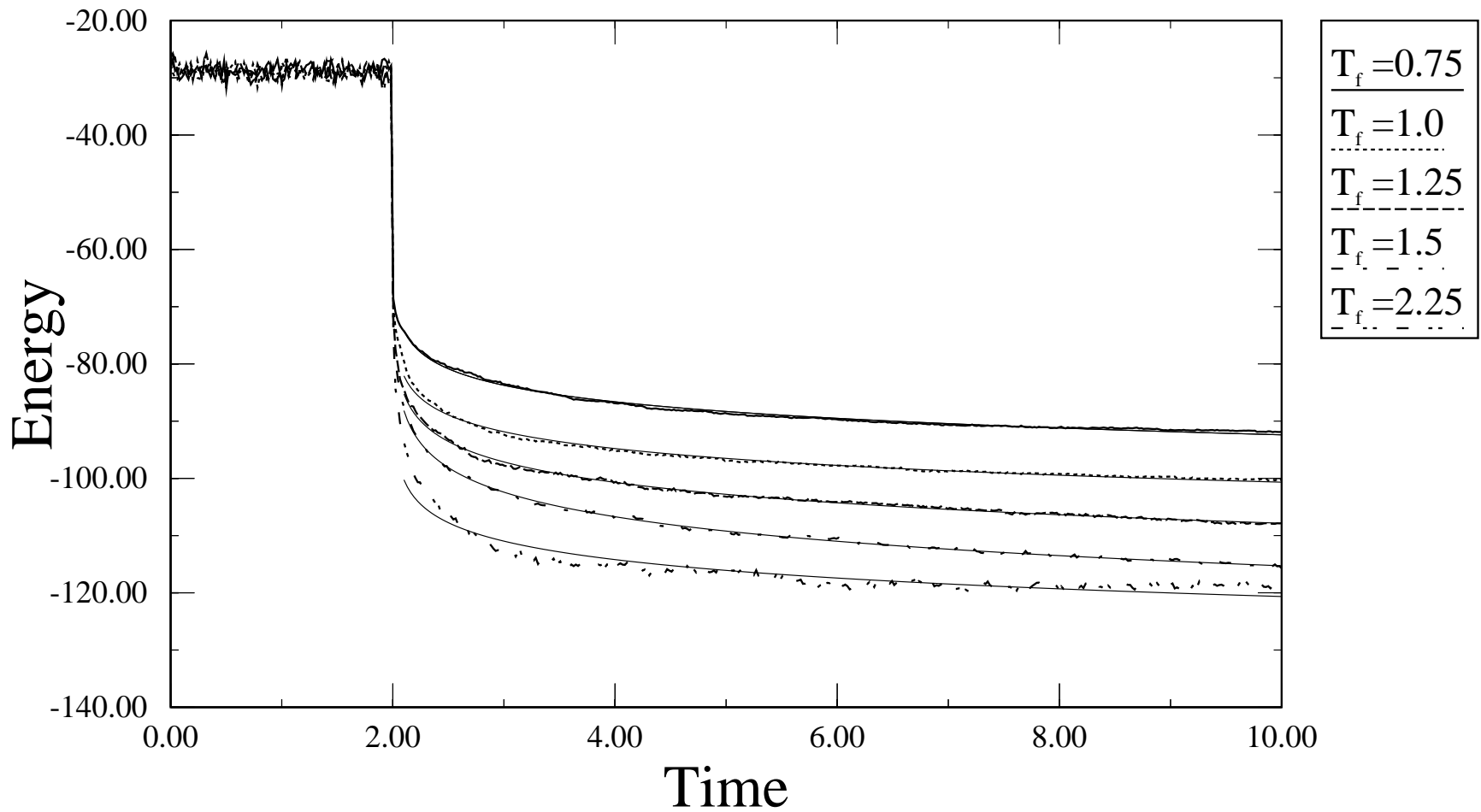
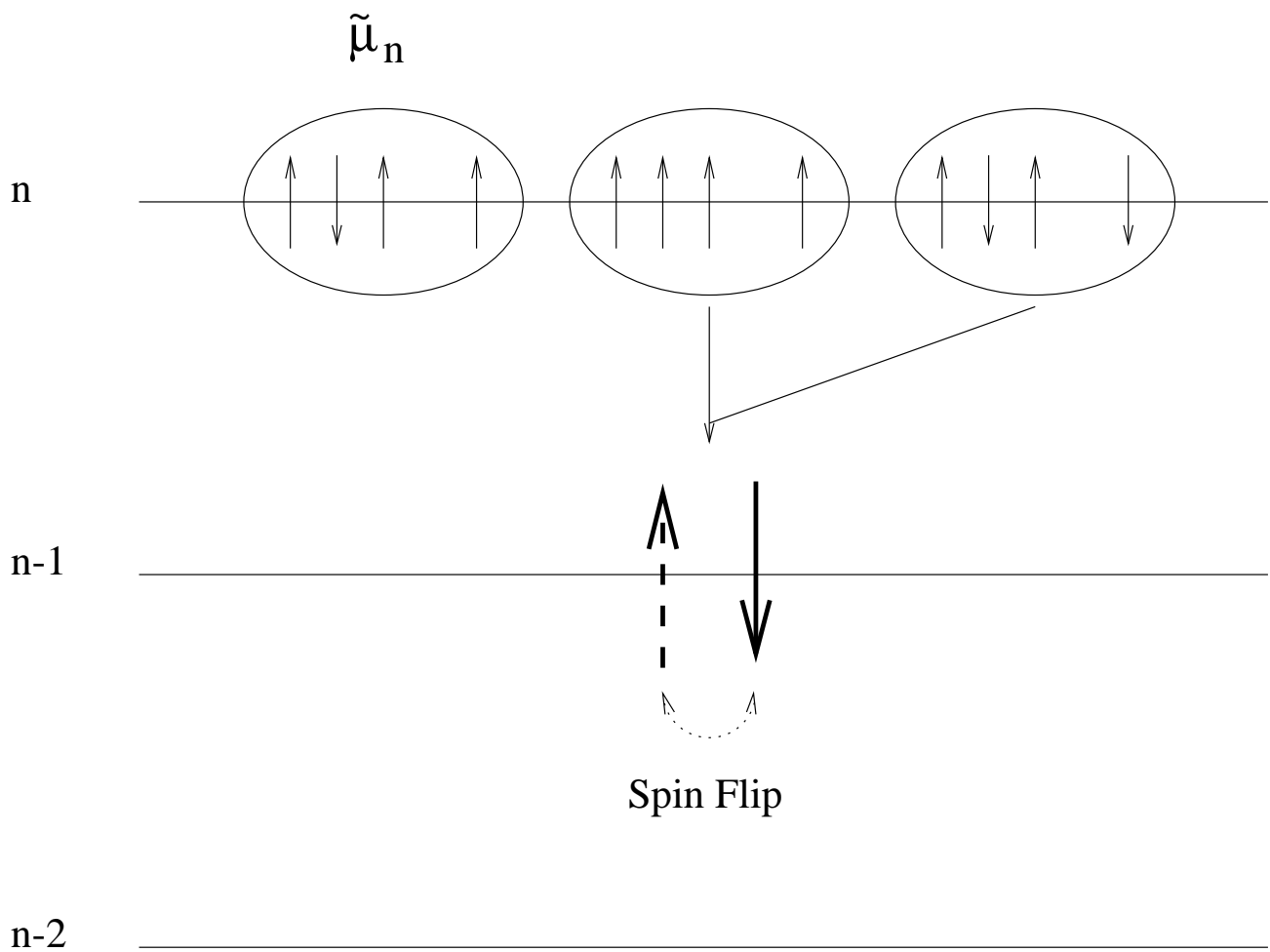
\end{document}